\documentclass[a4paper, twocolumn]{revtex4-2} 
\usepackage{color}
\usepackage{graphics}
\usepackage{natbib}
\usepackage{epsfig}
\begin{document}
	
\title{Large magnetodielectric coupling in the vicinity of metamagnetic transition in $6H-$perovskite Ba$_3$GdRu$_2$O$_9$}
	
	\author{S. Chhillar, K. Mukherjee and C. S. Yadav}
	\affiliation{ 
		School of Basic Sciences, Indian Institute of Technology Mandi, Kamand, Mandi-175075 (H.P.)
		India}

\begin{abstract}
The 6H-perovskites  Ba$_3$RRu$_2$O$_9$ (R = rare earth element) demonstrate the magnetodielectric (MD) coupling as a manifestation of $4d - 4f$ magnetic interactions. Here, we have reported a detailed study of the structural, magnetic, heat capacity, and MD properties of the 6H-perovskite  Ba$_3$GdRu$_2$O$_9$. The signature of long-range antiferromagnetic (AFM) ordering $ \sim$ 14.8 K ($T_N$) is evident from the magnetization and heat capacity studies. The $T_N$  shifts towards the lower temperature side, apart from splitting in two with the application of the magnetic field.  Field-dependent magnetization at 1.8 K shows three metamagnetic transitions with the opening of small hysteresis in different regions. A new transition at $T_1$ emerges after the onset of the first metamagnetic transition.  Complex magnetic behavior is observed in different magnetic field regions whereas  these field regions themselves vary with the temperature. Dielectric response recorded at zero and 80 kOe field exhibits the development of MD coupling well above $T_N$.  The MD coupling ($\sim$ 4.5 \% at 10 K) is enhanced by 25 \% as compared to the Dy counterpart. Effect of complex magnetic behavior is also conveyed in the MD results where the maximum value of MD coupling is observed in the vicinity of 10 K (onset of $T_1$) and near the second metamagnetic transition.  Our investigation suggests that both Gd and Ru moments align simultaneously at $T_N$. Short-range magnetic correlations are possibly responsible for MD coupling above $T_N$.  
\end{abstract}

\maketitle
\section{\label{sec:level1}Introduction:}

The study of MD effect, which enables the control of magnetization ({\it{M}}) and ferroelectric polarization ({\it{P}}) by electric field ({\it{E}}) and magnetic field ({\it{H}}) respectively, is one of the most promising field of condensed matter physics \cite{spaldin2019advances}. The study of MD coupling is crucial for technological applications as well as to understand the underlying physics \cite{maignan2018feNb,chaudhary2019Co4Ta2O9,lee2020Co4Ta2O9,Adroja2020_PRB_Ba2DyRuO6,Sanjay2017PRBR2BaNiO5,Surajit2020BaMnO3,Gastaldo2020NaMn7O12,Malik2020NdDyFeO3,AmelinTb2Ti2O7PRB}. In this respect, various compounds have been reported to show MD coupling viz. TbMnO$_3$ ($\sim$10\%) \cite{kimura2003magnetic}, EuTiO$_3$  ($\sim$7\%)\cite{katsufuji2001coupling}, Ba$_{0.5}$Sr$_{1.5}$Zn$_2$Fe$_{12}$O$_{22}$ ($\sim$13\%) \cite{kimura2005electric},  CaBaCo$_4$O$_7$ ($\sim$16\%)\cite{singh2012spin} and Tb$_2$BaNiO$_5$ ($\sim$18\%)\cite{upadhyay2017extraordinarily}.
Some of the promising candidate materials come from the 6H analogues family of BaRuO$_3$  $i.e.$  Ba$_3$RRu$_2$O$_9$ series  (where R = rare earth element) \cite{ling2016Ba3BiRu2O9,rath1994einkristall,samata2003synthesisBa3NaRu2O9,treiber1982Ba3BM2O9,huang2014BaBiIrRu2O9,middey2011glasslike,ziat2017Ba3MRu2O9InYLu,doi2001Ba3NdRu2O9,Doi2004Ba3MSb2O9,doi2001crystalBa3LnRu2O9LnCePrTb,doi2002Ba3MRu2O9,Doi2002GdHoYb,doi2002Ba3TbRu2O9,doi2004studiesBa3LnIr2O9LnLaNdSmYb,senn2013spin,basu2019HoTb,Basu2018,basu2020PRBHo,stitzer2002novel}.
In these compounds, the  R ion replaces one of the Ru lattice sites thus introducing new paths for the magnetic interactions. The size of the R ion decides the strength of Ru-Ru direct ferromagnetic (FM) interaction. The Ba$_3$MRu$_2$O$_9$ (M = Fe, Co, Ni, Cu, In) compounds are not suitable for dielectric applications \cite{ziat2017Ba3MRu2O9InYLu,doi2002Ba3MRu2O9} because of their semiconducting behavior. Substitution of the  rare earth ion at M site revives the insulating behavior \cite{basu2019HoTb}. Ru$_2$O$_9$ dimer is reported to have spin ½ moment distributed equally over both Ru atoms. This  Ru$_2$O$_9$ dimer forms a   frustrated quasi-two-dimensional triangular lattice and orders antiferromagnetically  \cite{ziat2017Ba3MRu2O9InYLu}. Intra-dimer interaction of Ru atoms in Ba$_3$RRu$_2$O$_9$  is still ambiguous \cite{doi2004studiesBa3LnIr2O9LnLaNdSmYb,basu2020PRBHo,senn2013spin}.  So far, MD coupling in Ba$_3$RRu$_2$O$_9$ series has been reported  for R =  Ho ($\sim$6 \%), Dy ($\sim$3.6 \%) Tb ($\sim$2 \%), Nd ($\sim$0.3 \%) and Sm ($\sim$0.05 \%) compounds\cite{basu2019HoTb,Basu2018,chhillar2021Dy}. Although the MD coupling in Ba$_3$HoRu$_2$O$_9$ compound sets $\sim$9.5 K, a significant amount of coupling is observed even at 50 K \cite{basu2019HoTb}.  Initially, magnetoelasticity was supposed to be responsible for it \cite{basu2019HoTb} but later it was found using neutron powder diffraction method that both Ru and  Ho orders simultaneously $\sim $50 K and spin reorientation occurs at 10 K \cite{basu2020PRBHo}. 
   
Here we have performed a detailed investigation of the magnetic and MD properties of  Ba$_3$GdRu$_2$O$_9$ compound. Preliminary magnetic and heat capacity analysis for this compound suggests the onset of AFM ordering $\sim$ 15 K \cite{Doi2002GdHoYb}. In addition to the highly insulating nature of this compound, the complex $4d - 4f$ magnetic interaction path is supposed to give large MD coupling. The Ba$_3$GdRu$_2$O$_9$ compound undergoes long-range AFM ordering $\sim $ 14.8 K ($T_N$) however short-range magnetic correlations are also present above $T_N$. Applied  magnetic field  forces the transition to shift towards low temperature side but no glassy behavior has been observed. Magnetic isotherms demonstrate that this compound shows the existence of meta-stable magnetic states which are manifested at different temperature and field regions.    HC data  demonstrate the splitting of  the peak anomaly at high fields similar to the magnetic measurement results. Entropy change corresponding to the transition at $T_N$ indicates that both Ru and Gd moments align simultaneously in the AFM structure. Dielectric response recorded at zero and 80 kOe field exhibits the development of  MD coupling well above $T_N$ due to the presence of short-range magnetic correlations.   The  MD coupling ($\sim$ -4.5 \% at 10 K) is enhanced by 25 \% as compared to the Dy counterpart and is the second largest among  Ba$_3$RRu$_2$O$_9$ (R = rare earth) family of compounds. This study points out the role of metamagnetic transition on the MD coupling.

\section{Experimental Details}
Solid state reaction method is used to synthesize the single-phase polycrystalline  compound Ba$_3$GdRu$_2$O$_9$ using BaCO$_3$, RuO$_2$ and Gd$_2$O$_3$ (Sigma-Aldrich, purity $ > 99.9\% $) precursors. Stoichiometric quantities were ground thoroughly using an agate mortar pestle and calcined at 900  $^o$C for 24 h followed by the heat treatments at 1200  $^o$C  (2$ \times$12 h) with several intermediate grindings and palletization. Final sintering was done at 1350 $^o$C.
Structural characterization was done using the Rigaku Smart Lab instrument with Cu-K$\alpha$ source. Rietveld refinement of obtained XRD data was performed using Fullprof Suite software  to check the single-phase formation of the compound.  SQUID magnetometer (Quantum Design Inc.) was used to perform the magnetization measurements. Heat capacity was measured using Physical Properties Measurement System (PPMS,  Quantum Design Inc.).  Hioki LCR meter with Cryonano Labs, India make sample holder integrated with PPMS was used for temperature and field dependent dielectric measurements. A rectangular sample was made as a parallel plate capacitor using silver paste on opposite surfaces and copper wires were used to make the contacts. The AC bias voltage of 1 V was used and data was collected at different frequencies by varying the temperature at zero and 80 kOe magnetic fields. Dielectric isotherms (for calculation of $\Delta \epsilon \%$) at different temperatures were taken by scanning the field from zero to 80 kOe and from 80 kOe to zero field at different frequencies.

\section{Results and Discussion}
\subsection{Structural Analysis}
\begin{figure}[hbt]
	\centering
	\includegraphics[width=\columnwidth]{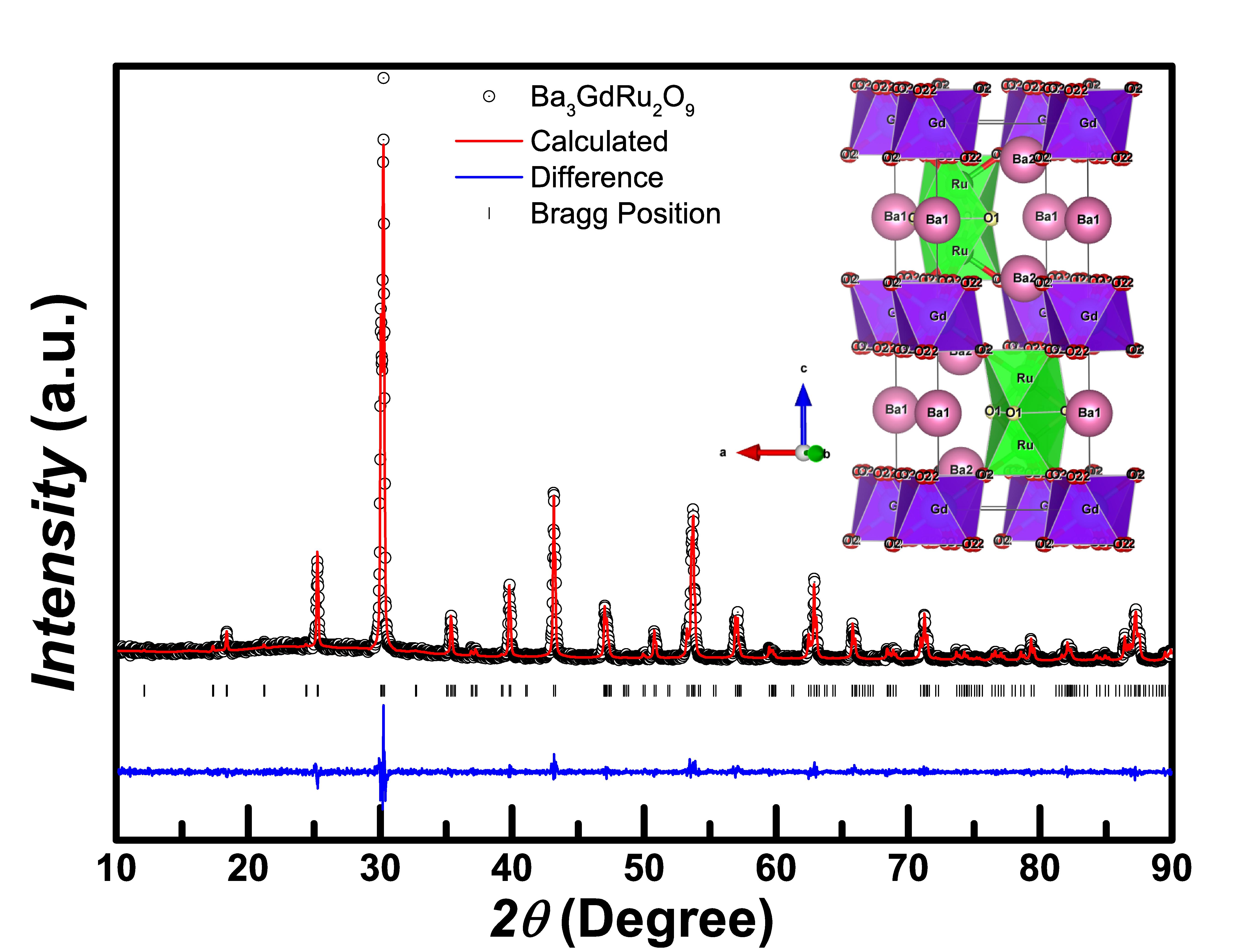}
	\caption{Rietveld refined x - ray diffraction data of Ba$_3$GdRu$_2$O$_9$. Inset shows the crystal structure.}
	\label{fig:xrd}
\end{figure}

Figure \ref{fig:xrd} displays the  Rietveld refined room temperature x - ray diffraction (XRD) pattern of polycrystalline sample Ba$_3$GdRu$_2$O$_9$ as well as the crystal structure.
The XRD pattern is refined using the six layered hexagonal ({\it{6H}}) crystal structure with space group $P6_3/mmc$ (No. 194).  Initial parameters for the refinement were taken from the literature\cite{Doi2002GdHoYb}. The Rietveld refinement confirms the single phase formation of the compound. Obtained lattice parameters are $a$ = $b$ =  5.9 \AA \ and $c$ = 14.6 \AA\ , which matches well with previous reports \cite{Doi2002GdHoYb}.
Crystal structure of Ba$_3$GdRu$_2$O$_9$ is represented in the inset of Fig. \ref{fig:xrd}. Detailed structural analysis of similar compound has been discussed earlier  \cite{chhillar2021Dy}.

\subsection{Magnetic Properties}

\begin{figure}[b]
	\centering
	\includegraphics[width=\columnwidth]{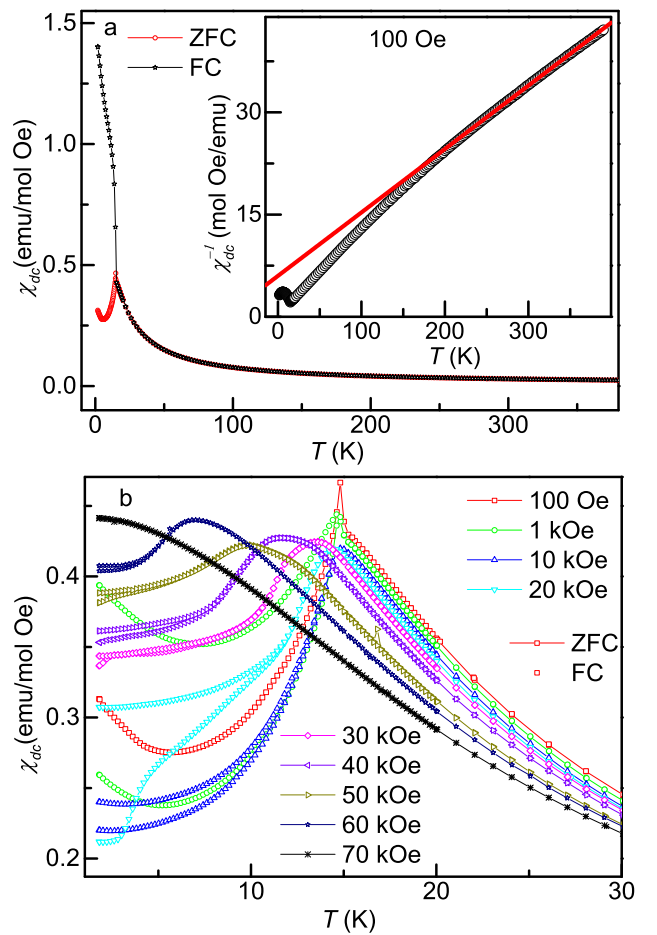}
	\caption{(a) The $dc$-magnetic susceptibility ($\chi_{dc}$)  at 100 Oe. Inset shows the Curie-Weiss fit (red line) (b) The $dc$-magnetic susceptibility ($\chi_{dc}$) at different fields. FC curve for 100 Oe is not shown in (b) part.}
	\label{fig:MT-MH}
\end{figure}

Fig. \ref{fig:MT-MH}(a) displays the $dc$-magnetization behavior of  Ba$_3$GdRu$_2$O$_9$ measured under the zero-field-cooled (ZFC) and field-cooled (FC) protocols at  100 Oe magnetic field.  Low field magnetic susceptibility  shows that this compound undergoes long-range AFM ordering $\sim$ 14.8 K ($T_N$).  Curie-Weiss fit  of  $dc$-susceptibility in the paramagnetic region using equation \ref{CW} is shown in the  inset of Fig. \ref{fig:MT-MH}(a). 

\begin{figure}[hb]
	\centering
	\includegraphics[width=\columnwidth]{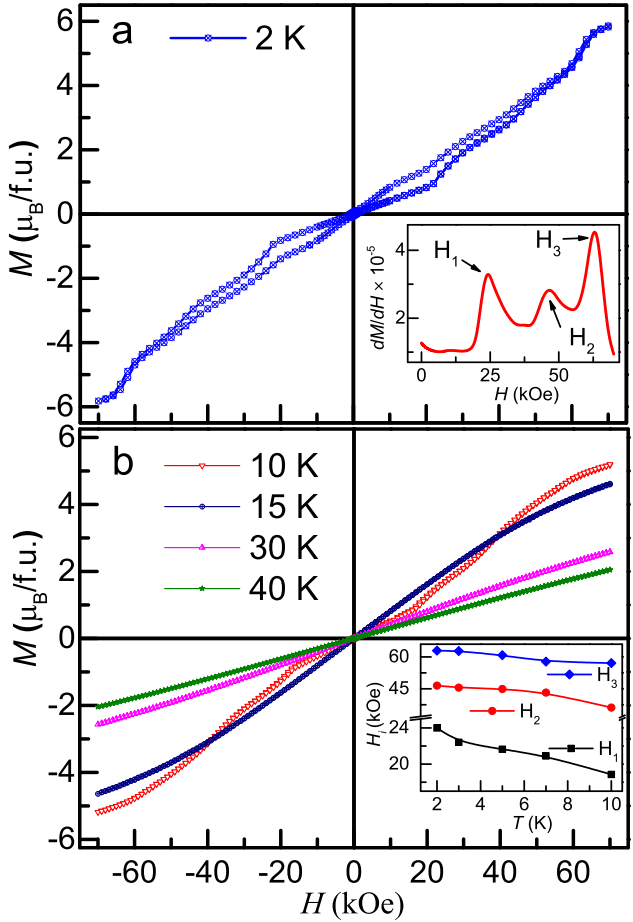}
	\caption{  (a) {\it{M-H}} curve at 2 K.  Inset shows the derivative of  {\it{M-H}} at 2 K (b) {\it{M-H}} curves at various temperatures. Inset shows the variation of metamagnetic transitions with temperature.}
	\label{fig:MH}
\end{figure}

\begin{eqnarray}\label{CW}
	\chi (T) = C/(T+\theta_W)\ where\ C = N_A \,\mu^{2}_{eff}/k_B 
\end{eqnarray}

Obtained values for Curie-Weiss temperature ($\theta_W$) and Curie constant ($C$) are $\approx$ -65 K and 16 $emu.K/mol.Oe$ respectively whereas the effective magnetic moment ($\mu_{eff}$) comes out to be  9.28 $\mu_B$. Assuming the free ion moment of Gd$^{3+}$ (7.94 $\mu_B$), the obtained value of $\mu_{eff}$ seems to have a contribution from the Ru atoms as well. Therefore the $\mu_{eff}$ per Ru atom is calculated using equation  $\mu_{cal} = \sqrt{(\mu_{Gd^{3+}})^2  + (\mu_{Ru^{4.5+}})^2   }$. The $\mu_{eff}$ per Ru atom at 100 Oe comes out to be 2.4 $\mu_B$  which is equal to the theoretical value of 2.4 $\mu_B$ (with $S = 1$ for Ru$^{4+}$ and $S = 3/2$ for Ru$^{5+}$). Data collected at larger magnetic fields (15 kOe, 35 kOe, and 52 kOe) yields the $\mu_{eff}$ = 8.5 $\mu_B$ when fitted with Curie -Weiss law in the temperature range of 140  – 300 K. The $\mu_{eff}$ per Ru atom, in this case, comes out to be 1.52 $\mu_B$ which is significantly less than that observed for lower fields and theoretical value. These results along with the deviation of inverse susceptibility curve  from linearity infer that the Ru atoms show weak short-range correlations at temperatures well above $T_N$. 

Further, increase in the field shifts the AFM ordering towards the lower temperature side. The peak width  also increases with the increase in the field (see Fig. \ref{fig:MT-MH}(b)). It appears that the application of high magnetic field broadens the transition peak as if two transitions are merged into one. Shift of the ordering temperature towards the lower side with the field is a characteristic feature of glassy systems. However, the $ac$-susceptibility measured at zero $dc$ field does not show any frequency dependent behavior. Thus it rules out the possibility of glassiness in the system.  Application of 1 kOe field reduces the bifurcation present in ZFC and FC curves which suggests the reduction of FM component present at 100 Oe field.  However, when 20 kOe field is applied, both ZFC and FC curves keep merged down to 11.5 K temperature and bifurcate below this temperature. The bifurcation temperature in ZFC and FC curves decreases to 5.4 K at 60 kOe while at 70 kOe field both ZFC and FC curves remain merged down to 2K. A broad shoulder like feature is also observed at 4.3 K temperature in the ZFC curve at 20 kOe field. This feature shrinks and shifts towards low temperature side with the application of high magnetic fields finally vanishing at 60 kOe field.   The susceptibility value at 1.8 K, decreases up to 20 kOe field and increases monotonically above this field. These results confirm that different field regions have different impacts on the magnetic state of the compound with new features appearing and disappearing with the application of the magnetic field.

The {\it{M(H)}}  isotherm collected at temperature 2 K, (Fig. \ref{fig:MH} (a)) shows three metamagnetic transitions at 24 kOe ({\it{H$_1$}}), 46 kOe ({\it{H$_2$}}) and 64 kOe ({\it{H$_3$}}), giving four regions of magnetic phases  (inset of Fig. \ref{fig:MH}(a)). The Ba$_3$HoRu$_2$O$_9$ compound also exhibit similar features attributed to spin-flip transition\cite{basu2020PRBHo}. Interestingly, hysteresis in {\it{M(H)}}  curve vanishes at 10 K (Fig. \ref{fig:MH}(b)).   The field, at which the metamagnetic transitions occurs, shifts towards lower field side and metamagnetic transitions also vanish above 10 K (Inset of Fig. \ref{fig:MH}(b)).  It is important to recall that $dc$-magnetization results (Fig.\,\ref{fig:MT-MH}) also show complex behavior in different field regions. Similarly, magnetic isotherms suggest that the magnetic state of the compound changes drastically in different field regions whereas these field regions themselves vary with the temperature. The {\it{M(H)}} curve for Ba$_3$GdRu$_2$O$_9$ remains non-linear even at 25 K temperature, suggesting the presence of magnetic correlations. The low value of   coercivity at 2 K which decreases with the increase in the temperature suggests a canted AFM  state at lower fields and low temperatures.

\subsection{Heat Capacity}
\begin{figure}[t]
\centering
 \includegraphics[width=\columnwidth]{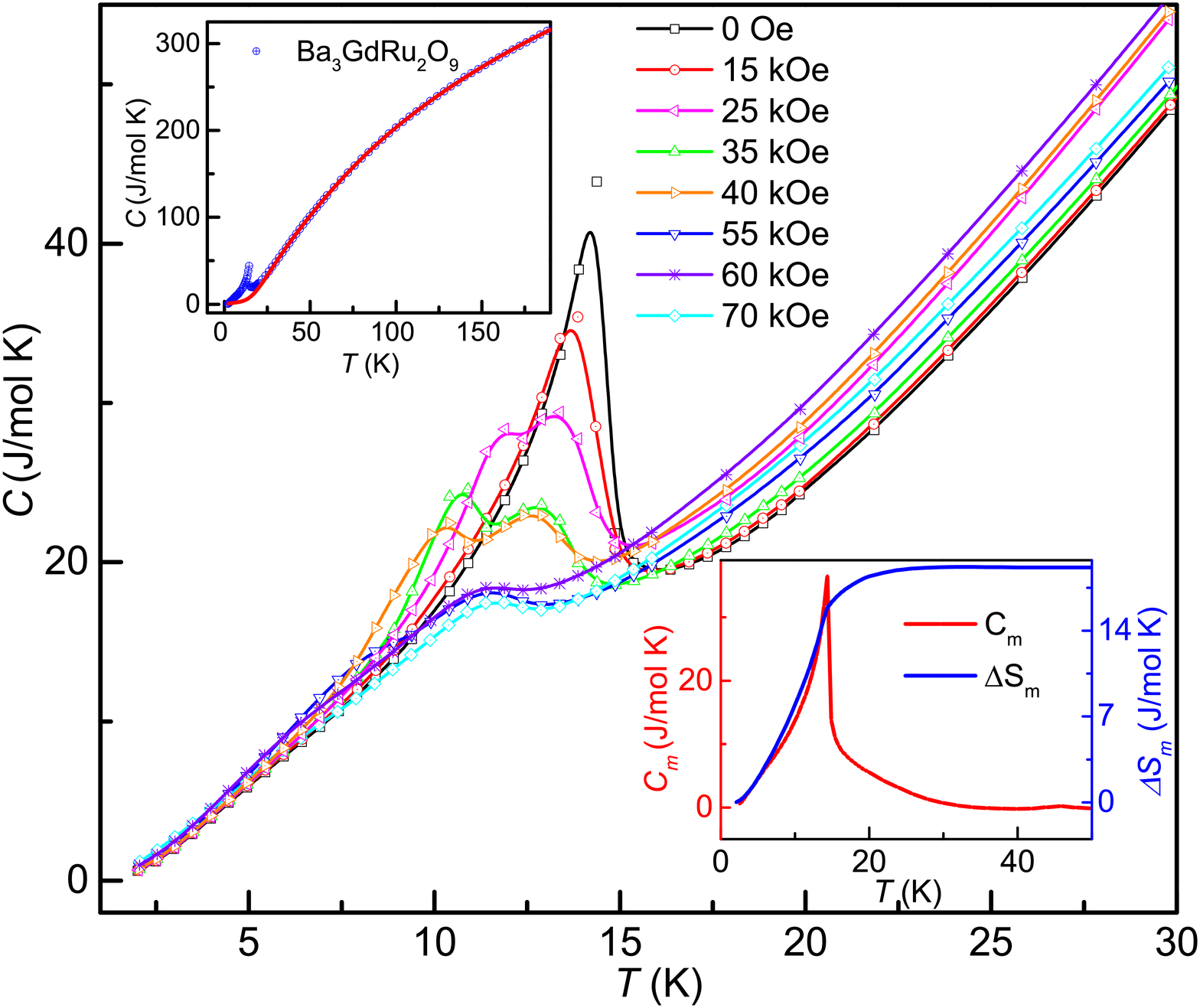}
\caption{Heat Capacity for Ba$_3$GdRu$_2$O$_9$ at different fields. Upper inset shows heat capacity at zero field where lattice part is fitted with 1 Debye and 2 Einstein model red line. Lower inset shows the magnetic heat capacity C$_{mag}$ (left) and corresponding entropy change (right).}
\label{fig:HC}
\end{figure}

 Zero field heat capacity (HC) data shows a clear anomaly $\sim$ 14.8 K  (see Fig. \ref{fig:HC}). This anomaly corresponds to the long-range AFM ordering temperature observed in the ZFC curve measured in 100 Oe fields.  The HC of a compound has contributions from magnetic, electronic, and lattice parts. HC data in the temperature range of 50 - 200 K was fitted in order to get the electronic and lattice contribution to the total HC (see upper inset of Fig. \ref{fig:HC}). We have used one Debye and two Einstein modes to get the lattice part of the heat capacity.  Lattice and electronic contributions to the HC were subtracted from the total HC in order to get the magnetic contribution ($C_{mag}$) to HC. The magnetic entropy change associated with the magnetic transition was calculated using equation, $\Delta S = \int{(\frac{\Delta C}{T})\Delta T}$  which came out to be 19.16 $Jmol^{-1}K^{-1}$ (see lower inset of Fig. \ref{fig:HC}). This value is larger than the entropy change corresponding to the ground state of  Gd$^{3+}$ ion $i.e.$ R ln8 = 17.4 $Jmol^{-1}K^{-1}$. Doi $et \, al$  found the entropy change to be 17.5 $Jmol^{-1}K^{-1}$ but that is not the case with our data \cite{Doi2002GdHoYb}. The reason behind this could be the different fitting models adopted. Excess entropy change might be due to the ordering of Ru${_2}^{4.5+}$ O$_9$ dimer which is having single unpaired electron \cite{Doi2002GdHoYb}.  These results indicate that both Ru and Gd moments order simultaneously in the AFM structure at $T_N$.

The temperature at which the anomaly is observed, shifts towards the lower side with the application of the magnetic field. Another anomaly appears at a magnetic field of 25 kOe.  With the further increase in the field, temperatures of both anomalies shift towards the lower temperature side along with the dilution of the anomaly peak. Weak features of both anomalies are observed even at a high field of 70 kOe. It is to be noted here that the application of magnetic field also broadened the transition in magnetization results as if there are two transitions present very closely. Now the HC results have clarified that those were indeed two transitions that were merged into one broad anomaly in the magnetization data. The Ba$_3$DyRu$_2$O$_9$ compound is also reported to have two anomalies in the zero field HC data which shifts towards low temperature side at high magnetic fields \cite{chhillar2021Dy}.

\subsection{Dielectric Analysis}

\begin{figure}[b]
	\centering
	\includegraphics[width=\columnwidth]{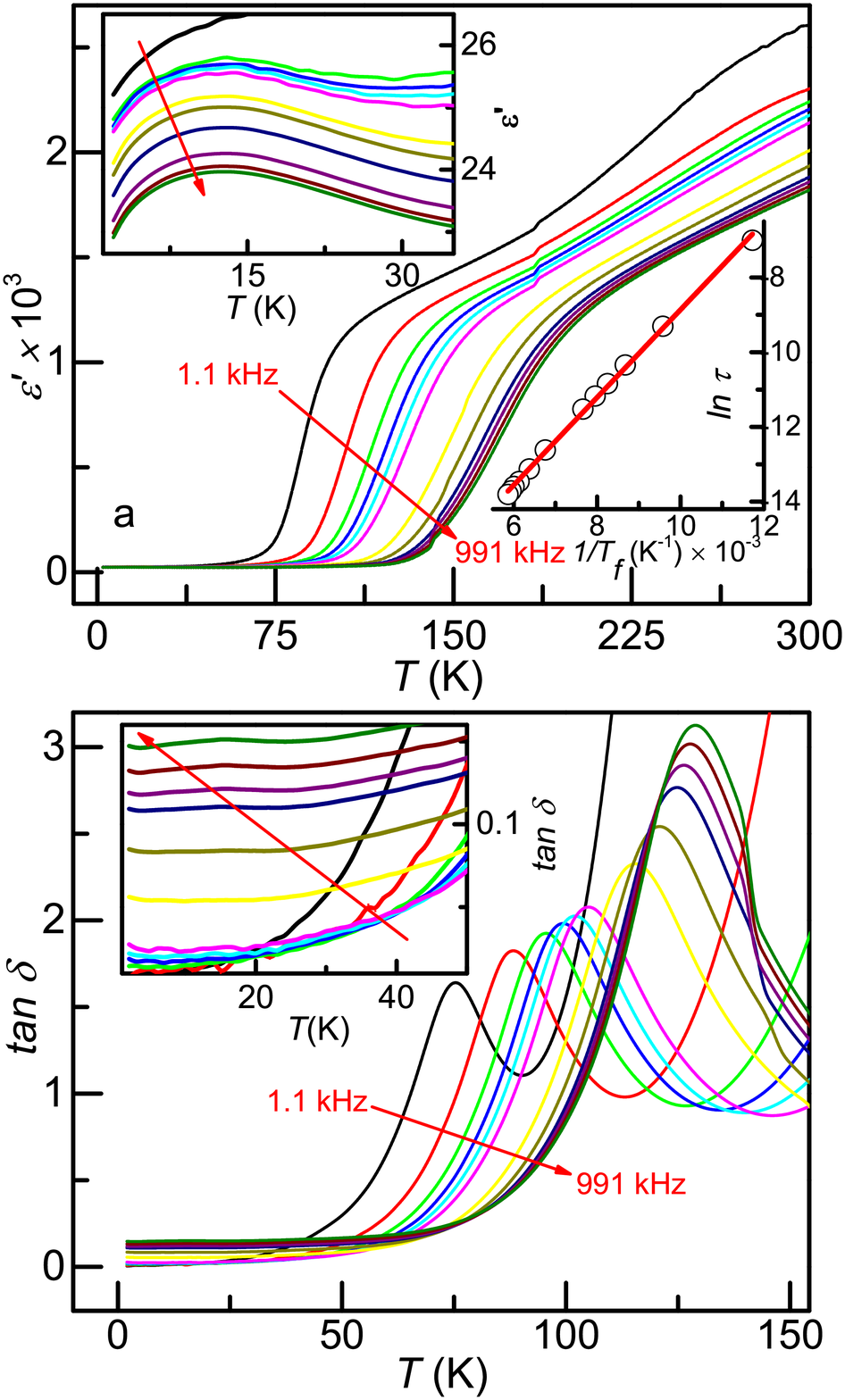}
	\caption{ Zero field data for Ba$_3$GdRu$_2$O$_9$ (a) Real part of dielectric constant $\epsilon'$ { $\textit{versus}$} T measured at different frequencies. Upper inset shows zoomed in temperature range of 2 - 35 K. Lower inset shows the Arrhenius fit (ln $\tau$ versus 1/T$_f$)for the relaxor behavior for T = 75 - 130 K (b) The derivative of the real part of dielectric constant (${d\epsilon'}/{dT}$) {$\textit{vesus}$} $T$. Inset shows the $tan\,\delta$ {$\textit{versus}$} $T$.}
	\label{fig:ET}
\end{figure} 
 
Fig.\,\ref{fig:ET} shows the temperature response of real part of dielectric constant ($\epsilon'$) and the loss factor ($tan\,\delta$)  at different frequencies measured at zero magnetic field. Value of  $\epsilon'$ decreases sharply below 75 K giving a broad cusp around 15 K which coincides with the AFM ordering temperature $T_N$ (see upper inset of Fig. \ref{fig:ET}(a)). This feature does not shift with the frequency indicating the magnetic origin of the anomaly.

\begin{figure}[ht]
	\centering
	\includegraphics[width=8cm,height=8cm]{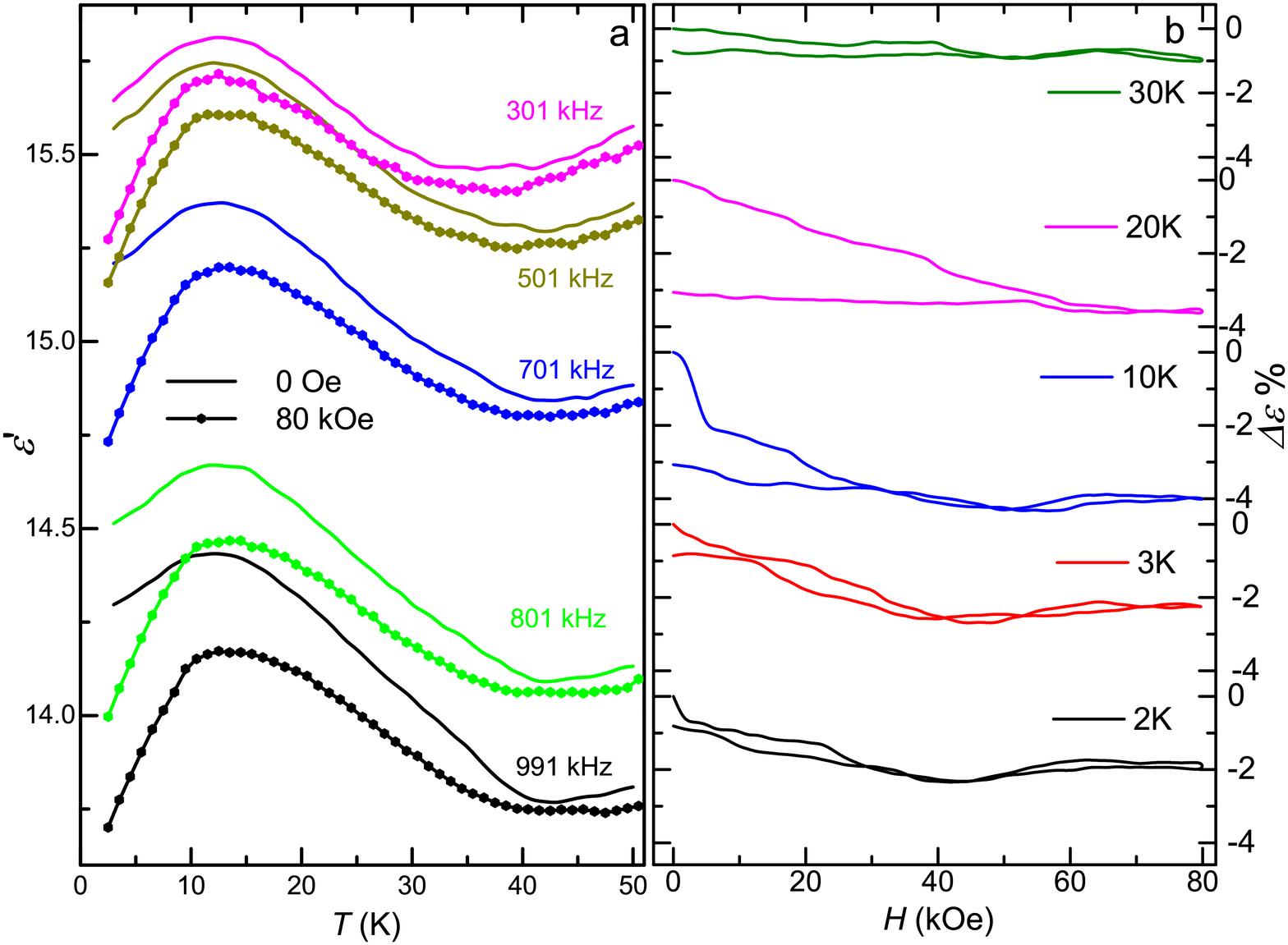}
	\caption{Temperature scan of real part of dielectric constant ($\epsilon'$) for different frequencies measured at zero and 80 kOe fields (b) Magnetic field dependence of magnetodielectric coupling ($\Delta\epsilon\%$)  measured at different temperatures.}
	\label{fig:MD}
\end{figure}

Value of $tan\,\delta$ is quite low ($\leq$ 0.2) below 50 K suggesting the absence of Maxwell-Wagner grain boundary effects in this temperature range (see Fig.\,\ref{fig:ET}(b)). The  $tan\,\delta$ curve also indicates the presence of frequency independent broad anomaly around $T_N$  as observed in the real part of dielectric constant. Loss part also shows frequency dependent anomaly with peak position shifting towards higher temperature side with the increase in frequency. Such a behavior is usually attributed to the thermally activated relaxor behavior \cite{Cross1987}. Peak position ($T_f$) is fitted with Arrhenius relation, $\tau = \tau_0\,exp(\frac{E_0}{k_BT})$, where $\tau_0$ and $E_0$ are relaxation time and activation energy respectively (see lower inset of Fig.\,\ref{fig:ET}(a)). The fitting resulted in the value of relaxation time and activation  energy to be $\approx$ 1.1 $\times$ 10$^{-9}  s$ and  $\approx$ 101 $eV$ respectively.  \\

To understand the repercussion of magnetic field on the dielectric response of the sample, $\epsilon' $ {\it{vs T}} is recorded at zero and 80 kOe magnetic fields and is plotted in the Fig.\,\ref{fig:MD}(a).
Both zero and 80 kOe field curves show a broad anomaly at $T_N$. It is interesting to note that the difference between zero and 80 kOe curves starts to develop around 40 K well above the long-range AFM transition at $T_N$. To calculate the MD coupling percentage, an isothermal field scan is performed at different temperatures. The MD coupling percentage is calculated using the equation;  $\Delta \epsilon \,\% =  100 \times  [\epsilon'(H) - \epsilon'(0)]/\epsilon'(0)$.\\
Negative MD coupling is observed throughout as shown in the  Fig.\,\ref{fig:MD}(b). The MD response at 2 K mimics the {\it{M(H)}} curve where hysteresis is present between the data taken while increasing and decreasing fields.  A maximum value of MD coupling ($\approx$ -4.3\%) is observed at 10 K temperature. The metamagnetic transitions also vanished above 10 K in  the MH data as discussed earlier. Significant MD coupling ($\sim$ -1\%) is still present at 30 K which is well above AFM ordering temperature ($T_N$). This might be due to the presence of short-range magnetic correlations above the magnetic ordering temperature. Ba$_3$RRu$_2$O$_9$ series compounds with R = Ho, Dy, Tb, Nd, Sm have been reported to show MD coupling of $\geq$6 \%, 3.6 \%, 2 \%, 0.3 \% and 0.05 \% respectively \cite{Basu2018,basu2019HoTb,chhillar2021Dy}. Thus the MD coupling in the current study is enhanced by 25 \% from the Dy counterpart which had shown the second largest value observed for Ba$_3$RRu$_2$O$_9$ series compounds.

\section{Discussion}

\begin{figure}[b]
	\centering
	\includegraphics[width=\columnwidth]{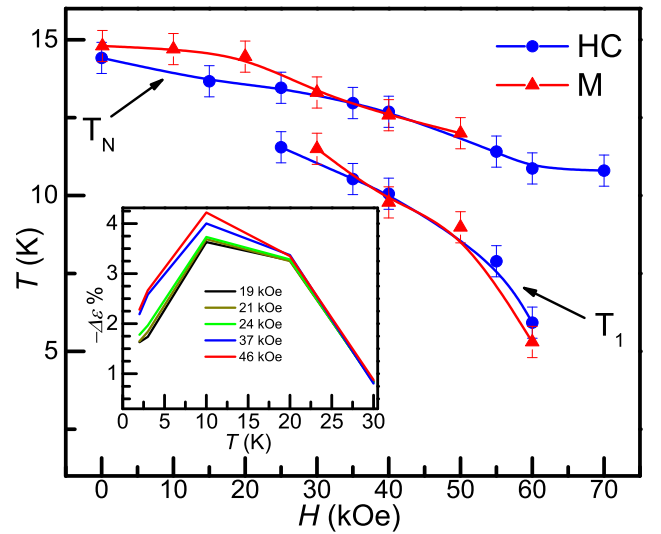}
	\caption{Variation of $T_1$ and $T_N$ with the field, obtained from magnetization results (M) and HC results. Inset shows the  $-\Delta\epsilon\%$ with temperature at different fields, obtained from MD results.}
	\label{fig:discussion}
\end{figure}

The Ba$_3$GdRu$_2$O$_9$ shows a complex magnetism with frustration parameter (f = $\theta_{CW}/T_{N})$ of $\sim$ 4. Low temperature magnetic isotherms confirm the presence of at least three metamagnetic transitions in the field range 0 - 70 kOe. Although there is a feeble signature of the metamagnetic states in the M(T) curves, the derivative of ZFC curves shows a new transition at $T_1$ below $T_N$. These results are further supported by the HC studies, where an additional peak is observed for the second metamagnetic state is also  evident for H $\sim$ H$_{1}$ ($\sim$24 kOe), besides the peak at T$_{N}$. We have plotted the $T_1$ and $T_N$ obtained from the magnetization and Heat capacity data in  Fig. \ref{fig:discussion}. As seen from the figure, transition at $T_1$ emerges in the already attained AFM state on the application of field larger than $H_1$. This behavior is rare but not impossible where a field induced transition has emerged at a different temperature. Interestingly, these transitions at $T_1$ and $T_N$ get suppressed for H $\geq$ 60 kOe and show a strong signal in the heat capacity measurement than in the magnetization studies. Similar observations of a weak transition in magnetization and distinctive behavior in HC have been reported for other 6H-perovskite  Ba$_3$DyRu$_2$O$_9$ compound \cite{chhillar2021Dy}. This behavior might be due to the fact  that contribution of Ru moment in overall moment of the compound is less so that ordering of Ru is not reflected clearly in the magnetization results. The 6H-perovskite Ba$_3$YbRu$_2$O$_9$, exhibits a distinct anomaly in the heat capacity above the AFM ordering, which is absent in the magnetization  data \cite{chhillar2021Dy}. We have shown the MD coupling parameter $\Delta\epsilon\%$ versus T measured at various magnetic field H in the inset of Fig. \ref{fig:discussion}  Interestingly, a maximum value of MD  coupling is also obtained in the vicinity of 10 K (onset of $T_1$) and 46 kOe (near $H_2$). Thus, it is believed the metamagnetic transition plays a crucial role on MD coupling.

\section{Conclusion}
In summary, we have performed a detailed study of structural, magnetic, and magnetodielectric properties of  Ba$_3$GdRu$_2$O$_9$ compound.  Although the Ba$_3$GdRu$_2$O$_9$ compound undergoes a long-range AFM transition  $\sim$ 14.8 K, short-range magnetic correlations are present above $T_N$.   Complex magnetic behavior is observed in different   magnetic field regions whereas  these field regions, themselves, vary with respect to the different temperature regions.   Dielectric response recorded at zero and 80 kOe field exhibits the development of MD coupling well above $T_N$. The  MD coupling ($\sim$ -4.5 \% at 10 K) is enhanced by 25 \% as compared to the Dy counterpart and is the second largest among  Ba$_3$RRu$_2$O$_9$ (R = rare earth) family of compounds. A maximum value of the MD coupling is observed in the vicinity of metamagnetic transition  ($H_2$).  Our investigation suggests that both Gd and Ru moments align simultaneously at $T_N$. Short-range magnetic correlations are possibly responsible for MD coupling above $T_N$. This study points towards the investigation of metamagnetic transition's effect on MD coupling not only in other members of the 6H-perovskites but also in other compounds which exhibit both phenomena concomitantly.\\
 
{\textbf{ACKNOWLEDGMENT:}} We acknowledge AMRC, IIT Mandi for the experimental facilities. S. Chhillar acknowledges UGC India for the financial support.  
 
\bibliographystyle{apsrev4-2}
\bibliography{GdArxiv}

\end{document}